\documentstyle[sprocl]{article}
\def\lsim{\mathrel{\rlap {\raise.5ex\hbox{$ < $}}
{\lower.5ex\hbox{$\sim$}}}}

\newcommand{\pr}{\paragraph{}}
\newcommand{\be}{\begin{equation}}
\newcommand{\ee}{\end{equation}}
\newcommand{\bea}{\begin{eqnarray}}
\newcommand{\nn}{\nonumber}
\newcommand{\eea}{\end{eqnarray}}
\newcommand{\nd}[1]{/\hspace{-0.6em} #1}
\newcommand{\nk}{\noindent}
\def\gappeq{\mathrel{\rlap {\raise.5ex\hbox{$>$}}
{\lower.5ex\hbox{$\sim$}}}}

\def\lappeq{\mathrel{\rlap{\raise.5ex\hbox{$<$}}
{\lower.5ex\hbox{$\sim$}}}}

\begin{document}
\begin{flushright}
CERN-TH/97-133 \\
CTP-TAMU-28/97 \\
ACT-10/97 \\
OUTP-97-29P \\
quant-ph/9706051 \\
\end{flushright}
\vspace{1cm}

\title{VACUUM FLUCTUATIONS AND DECOHERENCE 
IN MESOSCOPIC AND MICROSCOPIC SYSTEMS} 

\author{JOHN ELLIS}
\address{ Theory Division, CERN, CH-1211, Geneva, Switzerland }  

\author{N.E. MAVROMATOS}
\address{ P.P.A.R.C. Advanced Fellow, 
University of Oxford, Dept. of Physics
(Theoretical Physics),
1 Keble Road, Oxford OX1 3NP, United Kingdom} 

\author{D.V. NANOPOULOS~\footnote{Contribution to the {\it Symposium
on Flavor-Changing Neutral Currents: Present and Future Studies},
UCLA, Los Angeles, U.S.A., February 1997.}}
\address{ Center for
Theoretical Physics, Dept. of Physics,
Texas A \& M University, College Station, TX 77843-4242, USA, and \\
Astroparticle Physics Group, Houston
Advanced Research Center (HARC), The Mitchell Campus,
Woodlands, TX 77381, USA, and \\
Academy of Athens, Chair of Theoretical Physics, 
Division of Natural Sciences, 28 Panepistimiou Avenue, 
Athens 10679, Greece.} 

\maketitle\abstracts{We discuss recent experimental 
evidence of decoherence in a laboratory mesoscopic system
in a cavity,
from which we draw analogies with the decoherence that
we argue is induced by microscopic quantum-gravity fluctuations
in the space-time background.
We emphasize the parallel r\^oles played in both cases by 
dissipation through non-trivial vacuum 
fluctuations that trigger the
collapse of an initially coherent quantum state.
We review a phenomenological parametrization 
of possible effects of this kind in the
neutral kaon system, where they would induce CPT violation,
and describe some epxerimental tests.}

\section{Introduction and Summary}
\nk There is currently much debate whether microscopic black holes
induce quantum decoherence at a microscopic level. In particular,
it has been suggested~\cite{Hawking}
that Planck-scale black holes and other
topological fluctuations in the space-time background cause
a breakdown of the
conventional $S$-matrix description of asymptotic particle
scattering in local quantum field theory, which should be
replaced by a non-factorizable superscattering operator
$\nd{S}$ relating initial- and final-state density matrices:
\be
\rho_{out} = \nd{S} \rho_{in}
\label{hawk}
\ee
It has further been pointed out that, if this suggestion
is correct, there must be a modification of the usual
quantum-mechanical time evolution of the wave
function, taking the form of a modified Liouville equation
for the density matrix~\cite{ehns}:
\be
\partial _t \rho =  \frac{i}{\hbar} [\rho, H] + \nd{\delta H}\rho
\label{ehns}
\ee
The extra term in (\ref{ehns}) is of the form generally
encountered in the description of an open quantum-mechanical
system~\cite{markov,zurek}, 
in which observable degrees of freedom are coupled
to unobservable components which are effectively
integrated over, which may evolve from a pure state to a
mixed state with a corresponding increase in entropy. Any
such evolution entails a violation of CPT, though in a
different form from that sometimes proposed in the context
of conventional space-time quantum 
mechanics~\cite{dalitz}~\footnote{Proposals
for violations of CPT within the quantum-mechanical context of pure 
states have also been made recently
in the 
context of critical string theory~\cite{kostel} . In 
particular, the authors of ref.~\cite{kostel}
have proposed certain perturbative
backgrounds of critical   
strings that violate Lorentz invariance 
and hence CPT.}.
\pr
The necessity of a mixed-state description is generally accepted
in the presence of a macroscopic black hole, but is far from
being universally accepted in the case of microscopic virtual
black-hole fluctuations. In a research programme to elucidate this
question, we have been analyzing
the possibility of quantum decoherence in a non-critical
formulation of string theory \cite{EMN}, and indeed found 
an extra term in the quantum Liouville equation of the
form conjectured in (\ref{ehns}). 
In the absence of a satisfactory treatment of quantum gravity,
the results of our work can
only be regarded as indicative. However, we think that they
constitute interesting circumstantial evidence in favour of the
picture advanced previously~\cite{Hawking,ehns,EMN}, namely
that microscopic quantum fluctuations in the space-time background
may induce a loss of quantum coherence in apparently isolated
systems. Moreover, the magnitude of $\nd{\delta H}
\sim {\cal O}(\frac{E^2}{M_P})$, with 
$E$ a typical low-energy scale, that we find
is consistent with previous string estimates, and may not lie
many orders of magnitude beyond the reach of particle physics
experiments in the neutral kaon system~\cite{emncpt,ELMN,emncplear}, 
that are sensitive
to this form of decoherence and the related CPT violation~\footnote{We 
stress that the approach advocated in ref. \cite{ehns,EMN}
is different from the model for quantum measurement
proposed in ref. \cite{grw}, which involves
a reduction of the wavefunction
by occasional `hits' introduced
{\it ad hoc}. We prefer the motivation for
gravity-induced decoherence, which may be
mathematically controlled in the 
context of non-critical strings~\cite{EMN}.}.
\pr
In this talk we
review briefly the formalism we propose,
as well as the present experimental limits on
this form of CPT violation. Before doing so,
however, we first review
recent experimental results~\cite{brune} confirming
the r\^ole of the environment in inducing
decoherence in mesoscopic systems, which
confirm theoretical expectations for 
the r\^ole of the environment in the transition 
from the quantum to classical worlds~\cite{zurek},
and bear close analogy with the quantum-gravity
phenomena that we advocate.

\section{Schr\"odinger's Cat in the Laboratory} 
\nk We start by reviewing the experimental situation
concerning the preparation of `Schr\"odinger's cats'
in the laboratory, basing our
discussion on~\cite{brune}.
To understand the physics behind the construction,
we find it instructive to define 
exactly what these `beasts' are, and how they can 
be constructed experimentally. There is a huge literature 
in Quantum Optics on this, which originated from
special studies of the behaviour of atoms 
in electromagnetic cavities~\cite{haroche}.
Schr\"odinger's cat is prepared by letting 
an atom, which may be in a quantum superposition 
of two states $e,g$, pass through a cavity containing (quantized)
electromagnetic radiation. The coupling of 
the atom to coherent cavity modes of radiation
has been tested experimentally in recent years~\cite{rabiexp}.
This coupling manifests itself through the so-called
`vacuum Rabi splitting', i.e., a splitting of the 
spontaneous emission~\cite{rabi} or absorption~\cite{agar}
spectra of atoms inside cavities containing coherent 
electromagnetic radiation, as a result 
of the interaction of the atom with coherent cavity modes. 
\pr
It is instructive to review a quantum-mechanical 
derivation of the Rabi-splitting phenomenon.
We concentrate on the case of the absorption spectrum~\cite{agar},
which is technically simpler, and also more relevant for the 
experimental situation of ref.~\cite{brune}. Consider 
the case of a system of $N$ two-level atoms with frequency 
$\omega_0$ interacting with a {\it single}-mode radiation field 
of frequency $\omega$. The relevant quantum-mechanical 
Hamiltonian is:
\be
H=\hbar\omega_0\sum_{i} S_i^z + \hbar \omega a^\dagger a 
+ \sum_{i} (\hbar \lambda S_i^+ a + H.C.)
\label{hamrabi}
\ee
where $a^\dagger,a$ are the creation and annihilation operators
for the cavity radiation-field modes, 
$S_i^z$,$S_i^{\pm}$ are the usual spin-$\frac{1}{2}$ 
operators, and $\lambda$ is the atom-field coupling. 
The atom-field system is not an isolated system, and there is 
{\it dissipation} due to the interaction of the system with the 
surrounding world. One important source of dissipation is 
the leakage of photons from the cavity at some rate $\kappa$.
If the rate of dissipation is not too big,
a quantum coherent state can still be formed, which would allow 
the observation of the vacuum-field Rabi oscillations. 
The density matrix $\rho$ of the atom-field system obeys a Markov-type 
master equation for the evolution 
in time $t$~\cite{agar}:
\be
\partial_t \rho =-\frac{i}{\hbar} [H, \rho] - \kappa (a^\dagger
a \rho - 2 a \rho a^\dagger + \rho a^\dagger a )
\label{markovrabi}
\ee
This is exactly the form of equation proposed~\cite{ehns} as
an appropriate description of decoherence effects in quantum gravity.
\pr
The limit $\kappa << \lambda \sqrt{N}$ in (\ref{markovrabi}) guarantees
the possibility that a coherent quantum
state may be formed, i.e., this limit describes environments that 
are weakly coupled to the system, whose 
decoherence times (see below) are therefore very long. 
In this limit, one can concentrate on the off-diagonal elements
of the density matrix, and make the following 
`secular' approximation for their evolution~\cite{agar}:
\be
\partial _t \rho _{ij} =-\frac{i}{\hbar} 
(E_{i} - E_{j} )\rho_{ij} - \Gamma _{ij} \rho_{ij} 
\label{secular}
\ee
where $\Gamma _{ij}$ denotes the damping factor, related
to the weak coupling of the atom-field system 
to the environment.  The analysis of ref. \cite{agar}
pertained to the evaluation of the susceptibility tensor
of the system, $\chi_{\alpha\beta}$,
which can be calculated by considering 
its interaction with an external field of frequency $\Omega$. 
The absorption spectrum is proportional to 
${\rm Im}\chi (\Omega) $, which has the form
\bea
 &~&{\rm Im}\chi(\Omega) ={\rm cos}^2\theta
\frac{\Gamma _-/\pi}{\Gamma _-^2 + 
\{ \Omega - \omega_0 + \Delta/2 -\frac{1}{2}
(\Delta ^2 + 4 N \lambda ^2 )^{1/2}\}^2}
+ \nonumber \\
&~&{\rm sin}^2\theta \frac{\Gamma _+/\pi}{\Gamma _+^2 + 
\{ \Omega - \omega_0 + \Delta/2 + \frac{1}{2}
(\Delta ^2 + 4 N \lambda ^2 )^{1/2}\}^2}
\label{suscept}
\eea
with $\Delta \equiv \omega_0 - \omega$.
In the above expression, the factors $\Gamma _{\pm}$ 
represent the damping in the equation of motion for the 
off-diagonal element of the density matrix
$<\Psi_0|\rho|\Psi_{\pm}^{S,C}>$,
where the $\Psi_{\pm}$ are eigenfucntions of $H$, classified by the
eigenvalues 
of the operators $S^2$, and $S^z+a^\dagger a \equiv C$. In 
the $N$-atom case under study~\cite{agar},
$S=N/2$ and $C=1 -N/2$.
The expression (\ref{suscept}) summarizes the 
effect of Rabi vacuum splitting in the absorption spectra of atoms: 
there is a doublet structure of the absorption spectrum
with peaks at:
\be
\Omega = \omega _0 - \Delta/2 \pm \frac{1}{2}( \Delta ^2 + 
4 N \lambda ^2 )^{1/2}
\label{rabiabs}
\ee
For resonant cavities, the splitting occurs with equal weights
\be
  \Omega = \omega_0 \pm \lambda \sqrt{N} 
\label{rabisplitting}
\ee
Notice here the {\it enhancement} in the effect 
for multiatom systems $N >> 1$. 
The quantity  $2\lambda$ is called the `Rabi frequency'~\cite{rabi}. 
\pr
There have been simple experiments which have confirmed this
effect~\cite{rabiexp}, involving
beams of Rydberg atoms, resonantly 
coupled to superconducting cavities. 
The situation which is of interest
for the decoherence experiments of ref.~\cite{brune} 
involves atoms that are near resonance with the cavity.
In this case, $\Delta << \omega_0$ but
$\lambda ^2 N/|\Delta |^2 << 1$, so that 
(\ref{rabiabs}) yields two peaks that are characterized
by pure dispersive shifts $\propto \frac{1}{\Delta}$:
\be
    \Omega \simeq \omega_0 \pm  \frac{N\lambda ^2}{|\Delta|} 
+ {\cal O}(\Delta)
\label{dispersive}
\ee
which is the case in the SC experiment of ref.~\cite{brune}.
\pr
Another important issue, which has been used in ref.~\cite{brune},
is the dephasing of the atom as a result of the 
atom-field Rabi entanglement described above. 
To understand better the situation, we
discuss a more generic case, that of a three-state
atom, $f,e,g$, with energies $E_g > E_e > E_f$. 
Suppose one is interested in the transition $f \rightarrow e$ 
by absorption, in the presence of atoms in interaction with a cavity
mode. Calling $D_{ef}^+ \equiv |e><f|$,$D_{ef}^- \equiv |f><e|=
(D_{ef}^+)^\dagger$, we have the following effective Hamiltonian
for the transition $f \rightarrow f$~\cite{qnd}:
\be
 H_{eff}^{ef}=\hbar \omega_{eff}D_{ef}^+D_{ef}^-
\qquad ; \qquad \omega_{eff}=\omega_{ef} + \frac{\lambda ^2 n}{\Delta}
\label{eftr}
\ee
where the effective dfrequency is due to the dispersive
frequency shifts (\ref{dispersive}) 
of the Rabi effect, appropriate for near-resonant
atom-cavity-field systems. Here $n$ is the number of cavity 
photons~\footnote{The $\sqrt{n}$ scaling law for the Rabi splitting
(\ref{rabisplitting})
is also valid in the case of the interaction  
of a single atom $n$ cavity oscillator quanta, e.g., in a 
cohherent cavity mode.}.
\pr
Consider now an experiment to measure, say, the 
photon number $n$ in the cavity. The relevant probe
$P$ can be the above-described three-state atom, in a superposition
of $e$ and $f$ states. In this picture, the photon number $n$ is 
an eigenvalue of the cavity signal operator $a_s^\dagger a_s$, and
the
interaction Hamiltonian between atom and cavity then reads~\cite{qnd}:
\be
H_I =\frac{\hbar\lambda^2}{\Delta} a_s^\dagger a_s D_{ef}^+D_{ef}^-
\label{iham}
\ee
The probe observable is the atomic dipole 
operator:
\be
   A_P=\frac{1}{2i}(D_{ef}^+-D_{ef}^-)
\label{dipole}
\ee
whose Heisenberg evolution equation is
\be
i\hbar \frac{d}{dt}A_P = [A_P, H_{eff}^{ef}+H_I]
\label{evolution}
\ee
from which it is easily seen that in a time 
interval $t$ the phase of the probe changes
by an amount:
\be
  \Delta \phi = \omega_{ef}t + \frac{\lambda^2 n}{\Delta}t
\label{phase}
\ee
The case of interest for the experiment of ref.~\cite{brune}
is a two-state atom. The resulting phase shift
is obtained from (\ref{phase}) by setting 
$\omega_{ef} =0$. Thus, in the experiment of ref.~\cite{brune},
the phase entanglement due to the atom-field Rabi coupling 
is
\be
   \Delta \phi _R = \frac{\lambda ^2 n}{\Delta}t
\label{rabiphase}
\ee
for a near-resonance atom-field system, with small 
detuning $\Delta $. 
\pr
We are now well equipped in to review the 
experiment of ref.~\cite{brune} in which
a mesoscopic Schr\"odinger's cat was constructed,
and the associated decoherence. 
The experiment involves sending a Rubidium atom, 
consisting of two circular Rydberg states $e$ and $g$,
through a microwave cavity storing a small coherent field
$|\alpha>$. The coherent cavity mode is mesoscopic in the sense that
an average number of photons is of order ${\cal O}(10)$. 
The atom-cavity coupling is measured by the Rabi frequency
$2\lambda /2\pi=48kHz$. The condition for Rabi dispersive
shifts (\ref{dispersive}) is satisfied 
by having $\Delta/2\pi$ in the range $[70 , 800 ]kHz$.
\pr
The atom is prepared in the superposition of $e,g$ states, 
by means of a resonant microwave cavity $R_1$. 
Then it crosses the cavity $C$, which is coupled to a 
reservoir that dissipates its energy on
a characteristic time scale $T_r << 1.5$ ms. 
A number of photons varying 
from $0$ to $10$ is injected by a pulsed source into the cavity $C$.
The field in the cavity relaxes to vacuum,
{\it dissipating} via
leakage of photons through the cavity, 
during a time $T_r$, before being regenerated for the 
next atom. The experiment is at an effective 
temperature of $T=0.6K$, which is low enough that
thermal effects are small. 
After leaving $C$, the atom passes through a second
cavity $R_2$, identical to $R_1$. One then measures the 
probability of finding the atom in the state $g$, say. 
The decoherence time is then measured for various 
photon numbers. This enables one to test the 
theoretical predictions that decoherence
between two `pointer states' of a quantum superposition
occurs at a rate proportional to the square of the distance
between the states. 
\pr
Let us understand this point better. 
The coherent oscillator states, characterizing 
the cavity modes, constitute a pointer basis:
an oscillator in a coherent state is 
defined by the average number of oscillator quanta $n$:
$|\alpha >:~|\alpha=\sqrt{n}$. 
Then, consider the measurement of the above-described 
experiment, according to which there is only a phase entanglement
between the cavity and the atom. The combined 
atom-cavity (meter) system is originally in the state
\be
   |\Psi >=|e,\alpha e^{i\phi}>+ |g,e^{-i\phi}> 
\label{super}
\ee
where the dephasing depends on the atomic level:
$\phi \propto \lambda^2 t/\Delta$, according to (\ref{dispersive}),
(\ref{rabiphase}). 
Coupling the oscillator to a reservoir that damps its energy 
in a characteristic time scale $T_r$ produces decoherence,
which according to the general theory~\cite{ehns,milburn,qnd,markov}
occurs in a time scale inversely proportional to 
the square of the distance between the `pointer' states $D^2$:
\be
   t_{decoh} = \frac{2T_r}{D^2} 
\label{decoh}
\ee
In the set up of ref.~\cite{brune}, the distance $D$ is 
given by 
\be
   D=2\sqrt{n}{\rm sin}\phi \simeq 2n^{3/2}\frac{\lambda^2 t}{\Delta}   
\label{distance} 
\ee
for Rabi couplings $2\lambda$, such that $\lambda^2 t n << \Delta $. 
For mesoscopic systems, $n \sim 10$ $D > 1$, and hence decoherence 
occurs over a much shorter time scale than $T_r$. In particular,
for $\Delta /2\pi \sim 70kHz$, the decoherence time 
is $0.24 T_r$. 
\pr
This concludes our brief review of the construction of a Schr\"odinger's
cat, and the associated `measurement process'. 
Notice that the above construction is made in two stages:
first it involves an interaction of the atom with the 
cavity field, which results in a coherent state of the 
combined `atom + meter', and then {\it dissipation} is induced 
by coupling the cavity (measuring apparatus) to the environment,
which damps its energy, thereby inducing {\it decoherence}
in the `atom + meter' system. The important point to realize is
that the more macroscopic the cavity mode is, i.e., the higher 
the number of oscillator quanta, the shorter the decoherence
time is. This is exactly what was to be expected 
from the general theory~\cite{zurek,ehns,emohn,EMN}. 

\section{Quantum Gravity as an Environment, and the
Induced Collapse of Wave Functions}
\nk We now argue that a similar situation
characterizes quantum-gravity 
vacuum fluctuations. There is a striking analogy between the 
cavity vacuum and the quantum-gravity one,
with its virtual topological fluctuations 
in space time. 
\pr
The problem of the interaction of 
low-energy propagating matter 
with a dissipative quantum-gravity 
environment consisting of 
virtual wormholes~\cite{coleman} was studied in ref.~\cite{emohn}, from a
`pheonmenological'
view point. Coleman had argued that the wormhole state 
was likely to be a coherent state, and used this argument
to support the the vanishing of the 
cosmological constant. 
However, the coherence assumption was questioned later, and in view of 
our subsequent studies of the nature of the space-time 
foam in quaantum gravity and/or string theory
we expect this not to be the case. 
However, one can still model the interaction Hamiltonian 
between operators describing the low-energy probe $O_P$ and 
the wormhole state $|a>$ as~\cite{coleman}
\be
   H_I \propto O_P (a^\dagger + a ) 
\label{wormholes}
\ee
where $a^\dagger,a$ are creation and anihilation operators
for the wormhole state. In the example of ref.~\cite{emohn},
$O_P$ was taken to be a four-fermion effective interaction
\be
  O_P \propto {\cal O}(\frac{1}{m_P^2}) {\overline \psi}_1
\gamma^\mu \psi_1 {\overline \psi}_2 \gamma_\mu \psi_2
\label{fourfermi}
\ee
A low-energy observer has to average out the
unobservable wormhole 
effects, with the result that 
the low-energy probe $P$ becomes an {\it open} system. 
In ref.~\cite{emohn}, the simple case of a Gaussian 
distribution for the wormhole configurations 
was assumed, and the time scale of the 
induced decoherence
of the low-energy probe $P$ was estimated,
using the phenomenological 
equation for the density matrix suggested in ref.~\cite{ehns}, which was
characterized by probability 
and energy conservation of the probe. 
In view of our discussion in the previous section, 
this coupling may be considered as a 
coupling with only phase damping for the 
atom. Thus, a sort of Rabi vacuum effect 
appears, but of course the 
nature of the effect is not due to quantum 
electrodynamics, but due to quantum gravitational 
intreractions. The r\^ole of the cavity is played by
the whole universe, or rather by the microscopic
space-time foam~\cite{Hawking,ehns}. 
\pr
As shown in ref.~\cite{emohn}, the enhancement of
the effect for large numbers of atoms,
as seen above in our discussion of the simple 
Rabi vacuum~\cite{agar} (\ref{rabisplitting}) 
also characterizes the wormhole probe-P coupling.
The decoherence of the off-diagonal elements of 
the density matrix $\rho (x,x')$, in a `pointer' basis 
$|x>$, where $x$ is the center-of-mass location in space time 
of a system of $N$ particles is of the form~\cite{emohn}: 
\be
\rho (X',X, t) \sim \rho_0 (X',X, t) {\rm exp}[-ND(X'-X)^2t]
\label{decohmoha}
\ee
where $D$ represents the coupling of a single 
particle with a single coherent mode of the wormhole state, and is 
estimated to be of order $D \sim m^6/M_P^3$, in a four-dimensional 
space-time, for a particle of mass $m$, with $M_P$ the Planck 
mass. 
In the estimate (\ref{decohmoha}) a uniform density of 
wormholes of the order of one per Planck volume in space time
was assumed,
and all other interactions
of the microscopic particles among themselves have been ignored. 
{}From (\ref{decohmoha}) 
one can readily see the characteristic feature that the decoherence rate 
is propdorotional to the square of the distance between 
the pointer states (\ref{decoh}), which is a generic
feature of Markov-type decoherence~\cite{milburn}. 
\pr 
The wormhole model assumed 
that on the average energy and probability are conserved. 
This was also the case in the atom-cavity 
entanglement case considered above, where there was only
a phase entanglement/damping. 
Such entanglement is capable of producing 
decoherence by itself, as is clear from the analysis 
of ref.~\cite{emohn}. 
\pr
Although the mesoscopic atom + cavity system considered
in ref.~\cite{brune} is also exposed in such quantum-gravity
vacuum effects, they are of course much, much weaker than the 
conventional Rabi coupling, and negligible in the 
apparatus of ref.~\cite{brune}. 
However, as was demontrated
in ref.~\cite{emohn}, in the case of macroscopic systems
quantum-gravity effects could conceivably lead to
rapid collapse.
Unfortunately, it is not possible at present
to see such effects in such a cavity experiment.
However, experiments to look for macroscopic quantum-gravitational
decoherence may
become possible in the future,
especially in very cold environments, such as SQUIDs~\cite{emohn2} or those 
in which Bose-Einstein condensation has been observed~\cite{boseinst}. 
{}From our point of view, the experiments of ref.~\cite{brune},
although due to the conventional quantum field theory 
of Quantum Electrodynamics in a cavity,
are nevertheless fascinating, in that they constitute the first
experimental evidence for the
environmentally-induced
collapse of a coherent 
quantum superposition of states. 

\section{Decoherence in a String Approach to Quantum Gravity}
\nk We have argued in ref.~\cite{EMN} 
that in string quantum gravity there are inherently 
unobservable {\it delocalized} modes, carrying information, which 
fail to decouple from light states 
in the presence of singular space-time fluctuations. 
The effective theory of the light states which are 
measured by {\it local} scattering experiments can be  
described by a non-critical Liouville string~\cite{aben,EMN}.  
The zero mode of the Liouville filed in such a string 
theory is identified in ref.~\cite{EMN} 
with a target time variable. 
This results in an irreversible 
temporal evolution in target space,
with decoherence and associated entropy production, as we now review.
\pr
The effective low-energy theory 
density matrix is:
\begin{equation}
\tilde \rho (local, t) \, = \, \int d(delocal)\rho (local, delocal)
\label{intout}
\end{equation}
where ${\tilde \rho}$ denotes the low-energy density matrix, and
the $delocal$ states play a role analogous to those of the
unseen states $|B>_I$ inside the black-hole horizon
in the arguments of ref. \cite{Hawking}.
The integration over $delocal$ in (\ref{intout})
ensures that the reduced density matrix $\tilde \rho$ is {\it mixed} in
general, even if the full $\rho (local, delocal)$ is pure. We have
argued that $\tilde \rho$ obeys a modified quantum Liouville
equation of the form~\cite{EMN}
\begin{equation}
\partial _t {\tilde \rho} = i [{\tilde \rho}, H] + \nd{\delta H} {\tilde \rho} 
\qquad : \qquad 
\nd{\delta H} = -i \Sigma_{i,j} \beta^i G_{ij}[\, \, , g^j]
\label{stringmodliou}
\end{equation}
where $H$ is the usual light-particle Hamiltonian, the indices
$(i,j)$ label all possible microscopically-distinct string
background states with coordinate parameters $g^i$, and $G_{ij}$
is a metric in the space of such possible backgrounds~\cite{zam}. We argue
that these are not conformally invariant once one integrates out
the $delocal$ degrees of freedom, and the $\beta^i$ are the
corresponding renormalization functions. 
These are non-trivial
to the extent that back reaction of the light particles on the
background metric cannot be neglected. 
Equations of the form (\ref{stringmodliou})
are quite generic in the context of non-critical string 
theories~\cite{EMN,kogan}. We note further that the background 
fields $g^i$ must be quantized, as a result of the summation
over world-sheet 
topologies in the Liouville string~\cite{EMN}. 
\pr
There are general properties of the Liouville system that follow
from the renormalizability 
of the world-sheet $\sigma$-model theory~\cite{EMN}.
These include{ {\it energy conservation} on the average, 
and 
{\it probability } conservation. Specific 
properties of the renormalization group on the two-dimensional
world-sheet~\cite{zam} entail {\it monotonic} entropy
increase,
$\partial _t S \propto \beta^i G_{ij} \beta^j \ge 0$, 
leading to a microscopic arrow of time.
The modified quantum Liouville equation (\ref{stringmodliou}),
can be cast in a form similar to that of the Markov-type 
evolution (\ref{markovrabi}) of an open quantum-mechanical system:
\be 
      \partial_t \rho =i [ \rho, H] - \sum_{m} \{ B_m^\dagger B_m, \rho \}_{+}
+ 2 \sum_{m} B_m \rho B_m^\dagger  
\label{markovtype}
\ee
where the `environment' operators $B$ are appropriately-defined 
`square roots' of the various  partitions of the operator 
$\beta^i G_{ij} \dots g^j $~\cite{EMN}.  
\pr
The maximum magnitude of effect that we can imagine is
\begin{equation}
\nd{\delta H} \simeq H^2 / M_P
\label{order}
\end{equation}
which would be around $10^{-19} ... 10^{-20}$ GeV for a typical 
low-energy probe, such as the 
neutral kaon system. A contribution to the evolution rate equation
(\ref{stringmodliou}) of this order of magnitude would arise if there
were some Planck-scale interaction contributing an amplitude
$A \simeq 1/M_P^2$ and hence a rate $R \simeq 1/M_P^4$, to be
multiplied by a density $n \simeq L_P^{-3} \simeq M_P^3$, yielding the
overall factor of $ \simeq 1/M_P$ shown in 
(\ref{order})~\cite{emohn}. Such an 
estimate was found in a pilot study of a scalar field in a 
four-dimensional black-hole background~\cite{elizabeth},
and has also been found in a  
Liouville-string representation of Dirichlet membranes~\cite{emnd}.
\pr
The associated entropy production is a signature of
decoherence. Indeed, 
one can demonstrate in this approach
exponential decay in time of the off-diagonal
elements of the density matrix in the string theory space $|g^i>$.
Moreover, the Markov equation (\ref{markovtype}) 
implies~\cite{gisin} a stochastic equation of Ito-Langevin type 
for the state vector $|\Psi>$ corresponding to the density matrix 
$\rho (g^i, t)=Tr|\Psi><\Psi|$, 
\bea 
&~&      |d\Psi> =-\frac{i}{\hbar}H|\Psi> dt + \sum_m (<B_m^\dagger>_\Psi
B_m - \frac{1}{2}B_m^\dagger B_m - \nn \\
&~&\frac{1}{2} <B_m^\dagger >_\Psi 
<B_m>_\Psi) |\Psi> dt + \nn \\
&~& \sum_m (B_m - <B_m>_\Psi )|\Psi>d\xi_m 
\label{ito}
\eea
where the $d\xi_m$ are complex differential random matrices associated 
with Brownian processes. 
The advantage of the latter `state-vector' formalism is
that it allows a {\it localization} 
of the state vector in an appropriate `measurement' channel,
to be identified with a ground state of the string~\cite{mn}, 
as a result of the `dispersion-entropy minimization' theorem
of ref.~\cite{gisin}.

\section{Decoherence and CPT Violation}
\nk The non-unitary evolution characterising 
non-critical string theory 
manifests an arrow of time. 
Everyday experience tells us that an arrow of
time is present macroscopically: our bit (at least) of the
Universe is expanding, and we are all of us getting older. 
On the other hand, no such arrow
of time is visible in our accepted fundamental laws of physics: Quantum
Field Theory is invariant under CPT, and time $t$ is just a coordinate in
General Relativity - the motion of the Earth around its solar orbit could
be reversed with no apparent problem. On the other hand, an arrow of time
appears in thermodynamics via the second law, which states that entropy
increases monotonically. The arguments of the previous sections raise
again the possibility that this could have a microscopic origin.
\pr
It has been pointed out in ref.~\cite{wald} that a 
microscopic arrow of time must appear if pure
states evolve into mixed states as suggested above, in the sense that
the strong form of the CPT theorem must be violated. Suppose there is
some CPT symmetry transformation $\Theta$ which maps initial-state
density matrices into final-state density matrices:
\begin{equation}
\rho'_{out} \, = \, \Theta \rho_{in}
\label{inout}
\end{equation}
and correspondingly
\begin{equation}
\rho'_{in} \, = \, \Theta \rho_{out}
\label{outin}
\end{equation}
where 
\begin{equation}
\rho_{out} \, = \, \nd{S} \rho_{in}, \, 
\rho'_{out} \, = \, \nd{S} \rho'_{in}
\label{page}
\end{equation}
It is easy to deduce from these equations that $\nd{S}$
must have an inverse:
\begin{equation}
\nd{S}^{-1} \, = \, \Theta^{-1} \nd{S} \Theta^{-1}
\label{inverse}
\end{equation}
which cannot be true if pure states evolve into mixed states,
entropy increases monotonically and the density matrix
collapses.
\pr
Although there are many people in the quantum gravity
community who suspect that some modification of quantum mechanics
may be necessary so as to incorporate decoherence associated with
black holes, there is disagreement whether this is necessarily
accompanied by CPT violation. This division of opinion is exemplified
by the viewpoints of Hawking and Penrose in ref. \cite{hpbook}: Hawking
is very reluctant to give up CPT, whereas Penrose accepts it as a
likelihood. The formalism we have developed definitely points in
the latter direction.
\pr
An explicit example where all the above issues are realized 
has been given in ref. \cite{emnd}, and will not be repeated here.
Even if you do not follow all the arguments leading to the string version
of the modified Liouville equation,
the latter still provides an interesting phenomenological framework
in which one can parametrize possible decoherence and CPT-violating
effects with a view to the experimental tests in the neutral kaon
system, which are reviewed in the next section.

\section{Testing Quantum Mechanics and CPT in the Neutral Kaon System}
\nk The neutral
kaon system has an enviable track record as a probe of fundamental
physics, ranging from P violation (the $\tau$-$\theta$ puzzle) and
CP violation to the motivation for charm coming from the absence of
strangeness-changing transitions. It is also known to provide very 
elegant tests of quantum mechanics, and provides the most stringent
available test of CPT at the microscopic level.
The formalism of decoherence and 
related CPT violation developed
above can be applied to the neutral kaon system, and experimental upper
limits given on such effects.
\pr
In our approach, the quantum-mechanical evolution equation
is modified to become
\begin{equation}
\partial _t \rho = -i(H \rho - \rho H^+) + \nd{\delta H}\rho 
\label{nqmevol}
\end{equation}
where $H$ is the conventional quantum-mechanical Hamiltonian, and
we can parametrize the modification term $\nd{\delta H}$ as~\cite{ehns}
\begin{equation}
  {\nd h}_{\alpha\beta} =\left( \begin{array}{c}
 0  \qquad  0 \qquad 0 \qquad 0 \\
 0  \qquad  0 \qquad 0 \qquad 0 \\
 0  \qquad  0 \qquad -2\alpha  \qquad -2\beta \\
 0  \qquad  0 \qquad -2\beta \qquad -2\gamma \end{array}\right)
\label{deltah}
\end{equation}
where the indices $\alpha, \beta$ label Pauli matrices
$\sigma_{\alpha, \beta}$ in the $K_{1,2}$ basis, and we
have assumed that $\nd{\delta H}$ has $\Delta S = 0$. 
The three free parameters $\alpha,\beta,\gamma$
must obey the conditions
\begin{equation}
\alpha , \, \gamma \, > \, 0, \qquad \alpha \gamma \, > \, \beta^2
\label{positiv}
\end{equation}
stemming from the positivity oof the matrix $\rho$. 
\pr
It is easy to see that these parameters 
induce decoherence and violate CPT~\cite{emncpt}. 
Various observables sensitive to these parameters have
been discussed in the literature, including the following
asymmetries, which have already been used in experimental
probes of this formalism: the $2 \pi$ decay asymmetry
\begin{equation}
A_{2 \pi} \equiv \frac{\hbox{Tr}(O_{2 \pi}{\bar \rho}(t)) \, - \,
\hbox{Tr}(O_{2 \pi} \rho(t))}
{\hbox{Tr}(O_{2 \pi}{\bar \rho}(t)) \, + \hbox{Tr}(O_{2 \pi} \rho(t))}
\label{twopiasy}
\end{equation}
where $O_{2\pi}$ is an observable measuring the rate of $2\pi$ decay, and
$\rho, {\bar \rho}$ denote the density matrices of states
that are tagged initially as pure $K, {\bar K}$ respectively, 
and the double semileptonic decay asymmetry
\begin{equation}
A_{\Delta m}={R(K^0\to\pi^+)+R(\bar K^0\to\pi^-)-R(\bar K^0\to\pi^+)
-R(K^0\to\pi^-)\over R(K^0\to\pi^+)+R(\bar K^0\to\pi^-)+R(\bar
K^0\to\pi^+)
+R(K^0\to\pi^-)}
\label{deltamasy}
\end{equation}
in which various systematic effects cancel. 
The asymmetry $A_{2 \pi}$ is sensitive to
the presence of $\alpha, \beta$ and $\gamma$, whereas
$A_{\Delta m}$ is particularly sensitive to $\alpha$.
These and other measurements would enable the form of decohering CPT
violation
that we propose here to be distinguished in principle
from ``conventional" quantum-mechanical CPT violation~\cite{ELMN,HP}.
\pr
Together with the CPLEAR collaboration itself, we have published 
a joint analysis of CPLEAR data~\cite{emncplear}, 
constraining the CPT-violating
parameters $\alpha, \beta, \gamma$. The data
for $A_{2 \pi}$ and $A_{\Delta m}$ agree perfectly with a
conventional quantum-mechanical fit, and provide the
following upper limits when we impose
the positivity constraints (\ref{positiv}):
\begin{equation}
\alpha < 4.0 \times 10^{-17} \hbox{GeV}, \qquad 
\beta < 2.3 \times 10^{-19} \hbox{GeV}, \qquad 
\gamma < 3.7 \times 10^{-21} \hbox{GeV}
\label{bounds}
\end{equation}
We cannot help being impressed that these bounds are in the
ballpark of $m_K^2 / M_P$, which is the maximum 
magnitude that we could expect any such effect to have.

\section{Outlook}
\nk The experimental verification of environmentally-induced decoherence, 
observed in 
the mesosocopic atom $+$ cavity systems of ref.~\cite{brune},
has opened the way for an
understanding of the transition from the 
quantum to classical worlds, as anticipated 
by theorists for a long time~\cite{zurek,ehns,emohn,EMN}. 
This experiment, although based on conventional QED
environmentally-induced decoherence, is
important for its analogies 
with the decoherence that may 
result from quantum-gravity vacuum fluctuations. 
Simple laws of scaling with the number of 
microscopic constituents suggest
that couplings
between the quantum-gravity vacuum and low-energy probes
might allow an observable enhancement of the gravitional 
decohereing effects in macroscopic 
systems. If true, this
could open the way for an understanding of the nature of 
quantum space time. 
\pr
We have discussed the density-matrix formalism
of open systems, and
applied it specifically to the
neutral kaon system, which is believed to be the most sensitive 
probe of quantum mechanics to date. Our approach and formalism
can in principle be distinguished from others by measuring a number of
different $K, \bar K$ decay asymmetries~\cite{ELMN}. 
We can offer
our experimental colleagues no guarantee of success in such an
experimental programme. Nevertheless,
we think that the importance of the issues
discussed here motivate a new series of microscopic experiments~\cite{Dafne} 
to test quantum mechanics and CPT. 

\section*{Acknowledgments}
\pr
N.E.M. and D.V.N. wish to thank CERN Theory Division for the hospitality
during the last stages of this work. 
The work of D.V.N. is supported
in part by D.O.E. Grant
DEFG05-91-GR-40633.

\end{document}